# Precise Request Tracing and Performance Debugging for Multi-tier Services of Black Boxes

*Zhihong Zhang[1,2], Jianfeng Zhan[1], Yong Li[1], Lei Wang[1], Dan Meng[1], Bo Sang[1]*
[1]*Institute of Computing Technology, Chinese Academy of Sciences*
[2]*The Research Institution of China Mobile*
*Contact Person: jfzhan@ncic.ac.cn*

**Abstract**

*As more and more multi-tier services are developed from commercial components or heterogeneous middleware without the source code available, both developers and administrators need a precise request tracing tool to help understand and debug performance problems of large concurrent services of black boxes. Previous work fails to resolve this issue in several ways: they either accept the imprecision of probabilistic correlation methods, or rely on knowledge of protocols to isolate requests in pursuit of tracing accuracy.*

*This paper introduces a tool named PreciseTracer to help debug performance problems of multi-tier services of black boxes. Our contributions are two-fold: first, we propose a precise request tracing algorithm for multi-tier services of black boxes, which only uses application-independent knowledge; secondly, we present a component activity graph abstraction to represent causal paths of requests and facilitate end-to-end performance debugging. The low overhead and tolerance of noise make PreciseTracer a promising tracing tool for using on production systems.*

## 1. Introduction

As more and more multi-tier services are developed from commercial off-the-shelf components or heterogeneous middleware and deployed on data centers **without the source code available**, both developers and administrators need a precise request tracing tool to understand and debug performance problems of large concurrent services of **black boxes**.

Previous work fails to resolve this issue in several ways. Most related work requires obtaining the source code of applications, middleware, and relies upon the developer's instrumentation of applications or middleware, and thus they cannot be used for tracing requests of services of black boxes. For example, Magpie [1] requires the source code of applications and specific platforms with the appropriate logging capabilities [2]. The work [2] [3] of HP Labs accepts the imprecision of probabilistic correlation methods, uses the time series analysis method and offline infers causal paths from the logged messages of distributed systems of black boxes. Probably closest to our work is BorderPatrol [4], which relies on knowledge of protocols to isolate events or requests and proposes a precise request tracing method. When a multi-tier service is developed from commercial off-the-shelf components or heterogeneous middleware, BorderPatrol [4] requires writing many protocol processors and has to know more specialized knowledge than pure black-box approaches.

In this paper, we present a new precise request tracing method and make the following contributions:

(1) We design a novel precise tracing algorithm to deduce causal paths of requests from interaction activities of components of black boxes. Our algorithm only uses **application-independent knowledge**, such as timestamps, end-to-end communication channels and process contexts, which are available from the operating system.

(2) We present a component activity graph (CAG) abstraction to represent causal paths of requests and facilitate end-to-end performance debugging of a multi-tier service. Our experiments show we can pinpoint performance problems by using CAGs to calculate changes in observed latency percentages of components.

The paper is organized as follows: Section 2 formulates the problem; Section 3 explains the system observation, presentation and analysis; Section 4 introduces the precise tracing algorithm; Section 5 evaluates the system. Section 6 summarizes related work; Section 7 draws a conclusion.

## 2. Problem statement

We treat each component in a multi-tiers service as a black box. In Fig.1, we observe that a request will cause a series of *interaction activities in the operating system kernel* (in short, activities), e.g. sending or receiving messages. This happens under the contexts (process or kernel thread) of different components. We record an activity of sending a message as $S_{i,j}$, which means a process *i* sends a message to a process *j*. We record an activity of receiving a message as $R_{i,j}$, which means a process *j* receives a message from a process *i*.

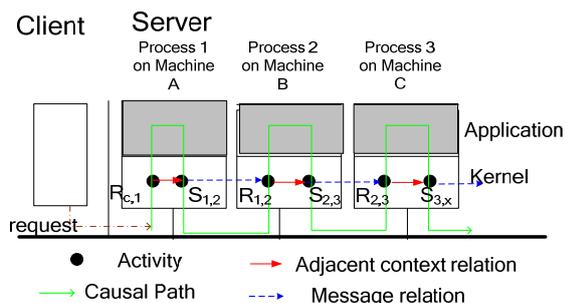

Fig.1. Activities with causal relations in the kernel.

When an individual request is serviced, a series of activities have causal relations or happened-before relationships, as defined by Lamport [7]. We define a sequence of activities with causal relations caused by an individual request as a *causal path*. For example in Fig 1, The activities sequence {$R_{c,1}$, $S_{1,2}$, $R_{1,2}$, $S_{2,3}$, $R_{2,3}$, $S_{3,x}$} constitutes a causal path. **For each individual request, there is a causal path**.

Our project develops a tracing tool to help developers and administrators in the following ways:
1) Precisely trace each request and correlate activities of components into causal paths.
2) Identify causal path patterns and obtain latency percentages of components for typical causal path patterns.
3) Debug performance problems of a multi-tier service with the help of 1) and 2).

The application limits of our system are as follows:
1) We treat each component in a multi-tiers service as a black box. We do not require the source code of the application, neither deploy the instrumented middleware, and neither have knowledge of high-level protocols used by the service, like http etc.
2) A single execution entity (process or kernel thread) of each component can only service one request in a certain period. For servicing each individual request, execution entities of components cooperate through sending or receiving messages using a reliable communication protocol, like TCP.

Though not all multi-tier services fall within the scope of our target applications, fortunately many popular services satisfy these assumptions. For example, our method can be used to analyze the concurrent servers following nine design patterns introduced in the book of Unix Network Programming by Stevens [5], including iteration model, concurrent process model and concurrent thread model.

## 3. The observation, presentation and analysis of system activities

### 3.1. The observation of system activities

We independently observe interaction activities of components of black boxes on different nodes. Concentrating on our main focus, we only concern about when servicing a request starts, finishes, and when components receive or send messages *within the confine of a data center*. Our concerned activities types include: BEGIN, END, SEND, and RECEIVE. SEND and RECEIVE activities are the ones of sending and receiving messages. A BEGIN activity marks the start of servicing a new request, while an END activity marks the end of servicing a request.

For each activity, we log four attributes: activity type, timestamp, context identifier and message identifier. For each activity, we use *(hostname, program name, ID of process, ID of thread)* tuple to describe its context identifier, and use *(IP of sender, port of sender, IP of receiver, port of receiver, message size)* tuple to describe its message identifier.

The instrumentation mechanism of our PreciseTracer depends on a open source software named SystemTap [8], which extends the capabilities of Kprobe[9]- the tracing tool on a single Linux node. Using SystemTap, we have written a module named TCP_TRACE, which obtains context information of processes and threads from the operating system kernel and inserts probe points into tcp_sendmsg and tcp_recvmsg functions of the kernel communication stack to log SEND and RECEIVE activities.

When an application sends or receives a message, a probe point will be trapped and an activity is logged. The original format of an activity produced by the TCP_TRACE is "timestamp hostname program_name ProcessID ThreadID SEND/RECEIVE sender_ip:port-receiver_ip:port message_size". PreciseTracer transforms the original format of an activity into more understandable n-ary tuples to describe context and message identifiers of activities. Distinguishing activity type is straightforward. SEND and RECEIVE activities are transformed directly. BEGIN or END activities are distinguished according to the ports of the communication channels. For example, the RECEIVE activity from a client to the web server's port 80 means the START of a request, and the SEND activity in the same connection with opposite direction means the STOP of a request in our algorithm.

## 3.2. The presentation and analysis of system activities

Formally, a causal path can be described as a directed acyclic graph G (V, E), where vertices V are activities set of components, and edges E represent causal relations between activities. We define this abstraction as *component activity graph* (*CAG*). For an individual request, a corresponding CAG represents all activities with causal relations in the life cycle of servicing an individual request.

CAGs include two kinds of relations: adjacent context relation and message relation. We formally define two relations based on happened-before relation [7], which is denoted as $\rightarrow$, as follows:

**Adjacent Context Relation**: if x and y are activities observed in the same context c (process or kernel thread) and caused by the same request r, and $x \rightarrow y$ holds true. If no activity z is observed in the same context, which satisfies the relations $x \rightarrow z$ and $z \rightarrow y$, then we can say an adjacent context relation exits between x and y, denoted as $x \xrightarrow{c} y$. So the adjacent context relation means that x has happened right before y in the same execution entity.

**Message Relation**: for servicing a request r, If x is a SEND activity which sends a message m, and y is a RECEIVE activity which receives the same message m, then we can say a message relation exists between x and y, denoted as $x \xrightarrow{m} y$. So the message relation means that x of sending message has happened right before y of receiving message in two different execution entities.

If there is an edge from activity x to activity y in a CAG, for which $x \xrightarrow{c} y$ or $x \xrightarrow{m} y$ holds true, then x is the parent of y.

In a CAG, every activity vertex must satisfy the property: each activity vertex has no more than two parents, and only a RECEIVE activity vertex could have two parents: one parent having adjacent context relation and another one having message relation with it.

Fig.1 shows a causal path for an individual request, where the red solid arrow represents adjacent context relation and the blue dashed arrow represents message relation.

For an individual request, it is clear that correlating a causal path is the course of building a CAG with the inputted interaction activities.

According to a CAG, we can calculate latencies of components in servicing an individual request. For example, for the request in Fig.1, the latency of process 2 is (t ($S_{2,3}$) - t ($R_{1,2}$)), and the latency of the interaction from process 1 to process 2 is (t ($R_{1,2}$) - t ($S_{1,2}$)), where t is the local timestamp of each activity. The latency of process 2 is accurate, since all timestamps are from the same node. The latency of the interaction from process 1 to process 2 is inaccurate, since we do not remedy the clock skew between two nodes.

We can classify CAGs into different *causal path patterns* according to the shapes of CAGs, since each CAG is a directed acyclic graph. Each causal path pattern is composed of a series of isomorphic CAGs, where similar vertices represent activities of the same type with the same context information.

For a causal path pattern, we aggregate and average n isomorphic CAGs to compute an *average causal path*. Furthermore, we obtain the latencies of components for an average causal path.

## 4. The precise tracing algorithm

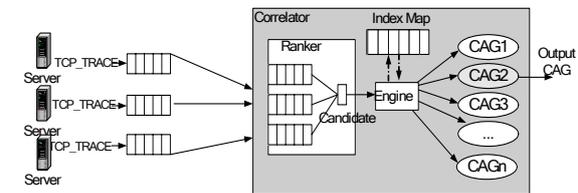

Fig.2. The architecture of PreciseTracer

Fig.2 shows the architecture of PreciseTracer, which is composed of two major components: TCP_Trace and Correlator.

Our precise tracing algorithm includes three main steps:

Step 1: When activities of components are logged by TCP_Trace on different nodes, they are sorted according to local timestamps in the first round.

Step 2: A module of Correlator, named *ranker*, is responsible for choosing candidate activities for composing CAGs.

Step 3: a module of Correlator, named *engine*, constructs CAGs from the outputs of the ranker, and then outputs CAGs.

Before we proceed to introduce the algorithm of the ranker, we explain how the engine stores unfinished CAGs. All unfinished CAGs are indexed with index map data structures. Index map maps a key to a value, and supports basic searching, inserting and deleting operations. One index map, named ***mmap***, is used to match message relations, and another one, named ***cmap***, is used to match adjacent context relations. For the mmap, the key is the message identifier of an activity, and the value of the mmap is an unmatched SEND activity with the same message identifier. The key in the cmap is the context identifier of an activity, and the value of the cmap is the latest activity with the same context identifier.

**4.1. Choosing candidate activity for composting CAGs**

We choose the minimal local timestamp of activities on different nodes as the initial time, and set a ***sliding time window*** for processing activities stream. Activities, logged on different nodes, will be fetched into the buffer of the ranker if their timestamps are within the sliding time window. The size of the sliding time window is independent of the clock skews, and it could be any value larger than 0, since our tracing algorithm does not depend on highly precise clock synchronization across distributed nodes.

The ranker puts activities into several different queues according to the IP addresses of the context identifiers of activities. Naturally, the activities in the same queue are sorted by the same local clock, so the ranker only need compare head activities of each queue and select candidate activities for composing CAGs according to the following rules:

**Rule 1**: If a head activity A in a queue has RECEIVE type and the ranker had found an activity X in the mmap, of which $X \xrightarrow{m} A$ holds true, then A is the candidate.

If the key is the message identifier of an activity A and the value of the mmap points to a SEND activity X with the same message identifier, we can say $X \xrightarrow{m} A$.

Rule 1 ensures that when a SEND activity has become a candidate and been delivered to the engine, the RECEIVE activity having message relation with it will also become a candidate once it becomes a head activity in its queue.

**Rule 2**: If no head activity is qualified with the rule 1, then the ranker compares the type of head activities of each queue according to the priority of BEGIN<SEND<END<RECEIVE<MAX and the head activity with lower priority is the candidate.

Rule 2 ensures that a SEND activity X always becomes candidate earlier than a RECEIVE activity A, if $X \xrightarrow{m} A$ holds true

After a candidate activity is chosen, it will be popped out from its queue and delivered to the engine, and the engine matches the candidate with an unfinished CAG. Then the element next to the popped candidate will become a new head activity in that queue. At the same time, the ranker will update the new minimal timestamp in the sliding time window and fetch new qualified activities into the buffer of the ranker in the new round.

**4.2. Constructing CAG**

The engine fetches a candidate, outputted by the ranker, and matches it with an unmatched CAG. Fig. 4 illustrates the pseudo-code of the correlation algorithm. In line 1, the engine iteratively fetches a candidate activity *current* by calling the function of the ranker, introduced in Section 4.1. From line 2-37, the engine parses and handles a ***current*** activity according to its activity type. Line 3-11 handles BEGIN and END activities. For BEGIN activity, a new CAG is created. For END activity, the construction of its matched CAG is finished.

______________________________________________
**Procedure** correlate {
1: **while** (current=ranker.rank ( )) {
2:   **switch** (current->get_activity_type ( )) {
3:    **case** BEGIN:
4:      create a CAG with ***current*** activity as its root.
5:    **case** END:
6:      find the matched parent where parent -<sup>c</sup>>current,
7:      **if** (the match is found) {
8:        add current into the matched CAG.
9:        add a context edge from parent to ***current.***
10:       output CAG.
11:     }
12:    **case** SEND:
13:      find matched parent_msg where parent_msg-<sup>c</sup>>current,
14:      **if** (the match is found) {
15:        If ( parent_msg.type==SEND )
16:          parent_msg.size += current.size.
17:        else {
18**:**       add current into the matched CAG.
19:          add a context edge from parent_msg to current.
20:     }
21:    }
22:    **case** RECEIVE:
23:      find matched parent_msg where parent_msg-<sub>m</sub>>current,
24:      **if** (the match is found) {
25:        parent_msg.size-=current.size.

```
26:    if (parent_msg.size ==0) {
27:        add current into matched CAG.
28:        add message edge from parent_msg to
               current.
29:        find matched parent_cntx where
               parent_cntx-^c >current,
30:        if (the match is found)
31:            if (parent_msg and parent_cntx are in the
                   same CAG)
32:                add a context edge from parent_cntx to
                       current.
33:      }
34:    }
35: }//switch
36:}//while
37}//correlate
```
Fig.3. The pseudo code of the algorithm.

Line 12-34 handle SEND and RECEIVE activities. The activities are inherently asymmetric between the sender and the receiver because of the underlying buffer sizes and delivery mechanism. So the match between SEND and RECEIVE activities is not always one to one, but n to n relation. Fig.4 shows a case that the sender consecutively sends a message in two parts and the receiver receives messages in three parts. Our algorithm correlates and merges these activities according to the message sizes in the message identifier tuples.

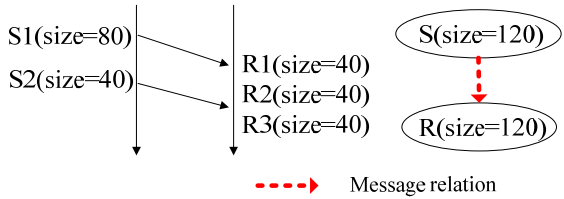

Fig.4. Merging multiple SEND and RECEIVE activities.

A situation may happen that an activity is wrongly correlated into two causal paths because of reusing threads in some concurrent programming paradigms. For example in thread-pool implementation, one thread may serve one request at a time. When the work is done, the thread is recycled into threads pool. Line 29-32 check if the two parents are in the same CAG. If the check returns true, the engine will add an edge of context relation, or else not.

### 4.3. Disturbance tolerance

In a clean environment without disturbance, Rule 1 and Rule 2 in Section 4.1 can produce correct causal paths. But in a practical environment, there may be two disturbances: noise activities and concurrency disturbance.

Noise activities are caused by other applications coexisting with the target service on the same computer. Their activities through the kernel's TCP stack will also be logged and gathered by our tool.

The ranker handles noise activities in two ways: 1) filters noise activities according to their attributes, including program name, IP and port. 2) If activities can not be filtered with the attributes, the ranker checks them with is_noise function. If true, the ranker will discard them. The pseudo algorithm of the is_noise function is illustrated in Fig. 5.

```
bool is_noise( Activity * E) {
return (( E->type==RECEIVE)&&(No matched SEND
activity X in the mmap with X-_m>E) && ( No matched
SEND activity Y in the buffer of ranker with Y-_m>E));
}
```
Fig.5. The pseudo code of is_noise ( ) function.

The second disturbance is called concurrency disturbance, which only exists in multi-processor computers. Fig.6-a illustrates a possible case of which two concurrent requests are serviced concurrently by two multi-processor computers and four activities are observed. $S^1_{1,2}$ means a SEND activity produces on the CPU1 of Node1, and $R^0_{1,2}$ is its matched RECEIVE activity produced on the CPU0 of Node2. When these four activities are fetched into the buffer of the ranker, they are put into two queues as shown in Figure 6-a. The head activities of both two queues are RECEIVE activities and block the matched SEND activities of each other. The ranker handles this case by swapping the head activity and the following activity in two queues. Figure 6-b illustrates our solution.

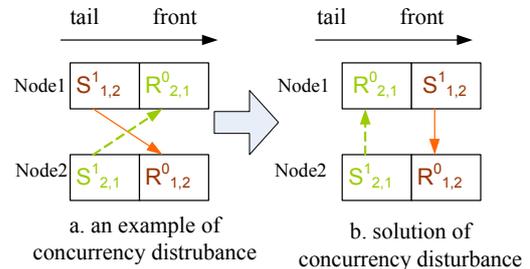

Fig.6. Example of concurrency disturbance.

### 4.4. The complexity of the algorithm

For a multi-tier service, the time complexity of our algorithm is approximately $O(g * p * \Delta n)$, where g measures the structure complexity of a service, p is the number of requests in the fixed duration, $\Delta n$ is the size of activities sequence per request in the sliding time window. Furthermore, the time complexity of our algorithm can be expressed as $O(g * n)$, where n is the size of activities sequence in the sliding time window. The space complexity of our algorithm is approximately $O(2g * p * \Delta n)$ or $O(2g * n)$.

## 5. Evaluation

### 5.1. Experimental environments and setup

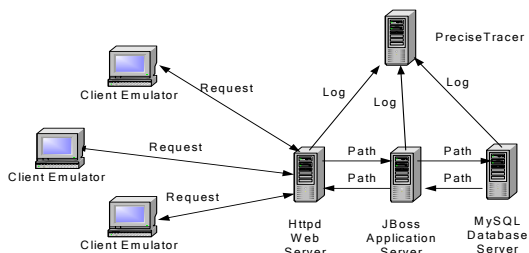

Fig.7. The deployment diagram of RUBiS.

We choose RUBiS [10] as the target application. Developed by Rice University, RUBiS is a three-tier auction site prototype modeled after eBay.com that is used to evaluate application servers' performance scalability.

The experiment platform is an 8-node Linux cluster connected by a 100Mbps Ethernet switch. Each node is a SMP with two PIII processors and 2G memory. Every node runs the Redhat Federo Core 6 Linux with the kprobe [9] feature enabled. The deployment of the RUBiS is shown in Fig. 7.

In the following experiments, *all experiments are done offline*. Client nodes emulate two kinds of workload: read only workload (Browse_only) and read_write mixed workload (Default). According to the user guide of RUBiS, every workload includes 3 stages: up ramp with the duration of 2 minutes 9 milliseconds, runtime session with the duration of 7 minutes 30 seconds 9 milliseconds, and down ramp with the duration of 1 minute 10 milliseconds.

### 5.2. Evaluating the accuracy of the algorithm

For RUBiS, we modify the code, tag and propagate a globally unique request ID with each request, and the following attributes are logged for the Apache web server, the JBoss Server and the MySql database, including (1) request ID, (2) the start time and end time of servicing a request; (3) ID of the process or thread.

At the same time, with only the application-independent knowledge, such as timestamps, end-to-end communication channels, we use PreciseTracer to obtain causal paths. From each causal path, we independently obtain information like (2) and (3).

If all attributes of a causal path are consistent with the ones obtained from the logs of RUBiS, we confirm that the causal path is correct. So we define the path accuracy as follows:

Path accuracy = correct paths/ all logged requests.

We test the accuracy of our algorithm for both Browse_Only and Default workload of RUBiS in many experiments with different configurations: (1) concurrent clients increase from 100 to 1000, and at the same time the size of the sliding time window varies from 1 millisecond to 10 seconds; (2) the clock skew changes from 1 millisecond to 500 milliseconds. (3) We run rlogin, ssh and MySQL client to produce noise activities.

In all these tests, the accuracy of our algorithm is 100% with no false positive and no false negative. This is because all activities of components are logged using SystemTap[8], so our algorithm correlates all activities into causal paths. However, for network congestion, the loss of activities may happen, though we have not observed up to now. The loss of activities will result in deformed CAGs. When the possibility of loss of activities is low, we can distinguish normal CAGs from deformed CAGs according to the difference of quantities.

### 5.3. Evaluating the efficiency of the algorithm

**5.3.1. Evaluating the complexity.** We set the sliding time window as 10 milliseconds. When concurrent clients vary from 100 to 1000, we record the number of serviced requests and the correlation time. For different numbers of concurrent clients, the test duration is fixed for the Browse_Only workload, defined in Section 5.1.

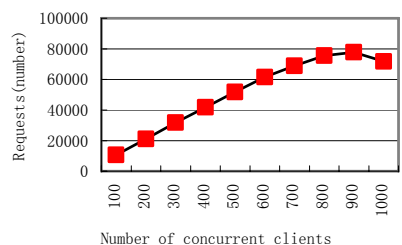

Fig.8. Requests v.s. concurrent clients.

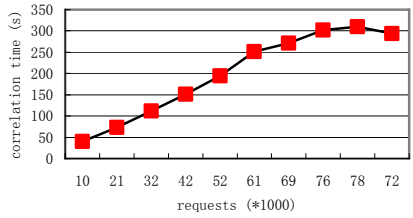

Fig.9. Correlation time v.s. requests. The unit in x-axis is 1000 requests.

From Fig.8 and Fig.9, we observe that the number of requests is linear in the number of concurrent clients and the correlation time is linear in the number of serviced requests in the fixed test duration before RUBiS is saturated at the point of 800 clients.

In Section 4.4, we conclude that the time complexity of our algorithm is approximately $O(g * p * \Delta n)$. Our experiment result in Fig.9 is consistent with the analysis, since g is a constant for RUBiS and △n is unchanged in the fixed sliding time window, so the correlation time is linear in the number of requests in the fixed test duration before RUBiS is saturated.

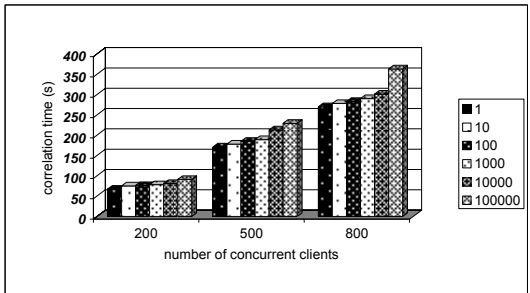

Fig.10. Correlation time v.s. sliding time window.

Fig. 10 shows the effect of the size of the sliding time window on the correlation time for different numbers of concurrent clients. Our analysis in Section 4.4 shows that, when the number of requests in the fixed duration is unchanged, the time complexity of the algorithm is linear in the size of △n for RUBiS. △n is the size of activities sequence per request in the sliding time window, which is determined by the size of the sliding time window.

Fig. 11 shows the effect of the size of the sliding time window on the memory consumed by the Correlator for different concurrent clients. When the size of the sliding time window increases from 10 seconds to 100 seconds, the number of logged activities, fetched to the buffer of the Correlator, increases dramatically and results in the dramatic increase of the consumed memory.

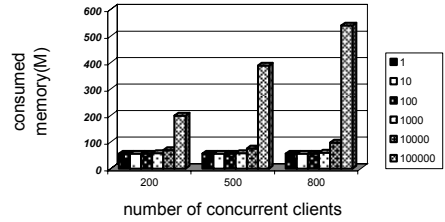

Fig.11. The memory consumed by the Correlator.

**5.3.2. Evaluating the overhead.** We compare the throughput and average response time of RUBiS for the Browse_Only workload when the instrumentation mechanism of PreciseTracer is enabled and disabled.

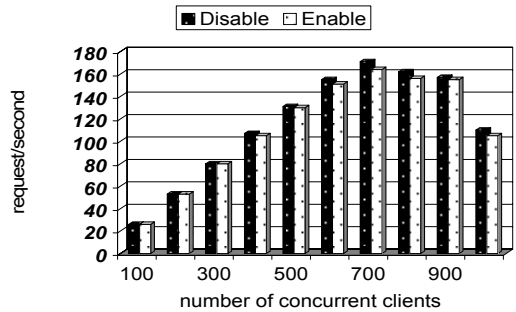

Fig.12. The effect on the throughput of RUBiS.

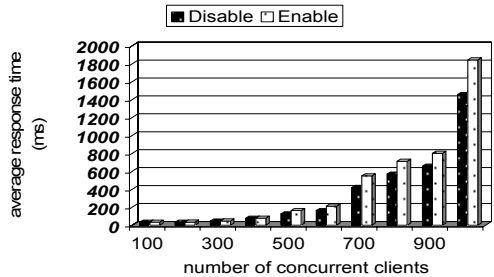

Fig.13. The effect on the average response time.

In Fig.12 and Fig.13, we observe that when the numbers of concurrent clients are less than 500, PreciseTracer has little effect on the throughput and average response time of RUBiS. As the number of concurrent clients increases, PreciseTracer has small effect on both of them. According to our statistics, the max overhead in terms of the increase of throughput is 3.7%, and the max overhead in terms of the increase of average response time is less than 30%.

**5.3.3. Evaluating noise tolerance.** We test the performance of PreciseTracer in an environment with the noise activities produced by rlogin, ssh and MySQL client. We set the size of the sliding time window as 2 milliseconds. In our algorithm, noise activities produced by rlogin and ssh can be filtered by the program name attribute, but noise activities produced by MySQL client cannot be filtered by the program name attribute, since MySQL client shares the same database with RUBiS. Our algorithm can discard these noise activities. Fig.14 measures the effects of noise activities on the correlation time when about 200K noise activities related with MySQL Database are produced in the fixed duration for the Browse_Only workload. We observe that our algorithm is effective in tolerating noise.

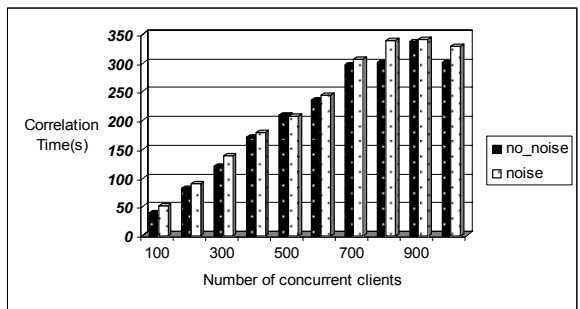

Fig.14. The overhead of noise tolerance.

### 5.4. Identifying performance bottleneck

**5.4.1. Misconfiguration shooting.** For the experiments in Section 5.3.2, we observe that when the number of concurrent clients increases from 700 to 800, the throughput of RUBiS decreases, while the average response time increases. An interesting question is what is the wrong with RUBiS?

Generally, we will observe the resource utilization rate of each tier and the metrics of quality of service to pinpoint the bottlenecks. Using the monitoring tool of RUBiS, we notice that the CPU usage of the each node is less than 80% and the I/O usage rate is not high. Obviously, the traditional method does not help.

To answer this question, we use our tool to analyze the most frequent request ViewItem for RUBiS, compute the average causal path and visualize the view of *latency percentages of components*. We identify the problem quickly.

From Figure 15, we observe that when the number of concurrent clients increases from 500 to more, the latency percentage of httpd2Java from first tier to second tier changes dramatically, and the value is 46%, 80%, 71% and 60% respectively for 500, 600, 700 and 800 concurrent clients. In Fig.15, the latency percentage of httpd2Java is 46% for 500 clients, which means that the processing time of the interaction from httpd to Java takes up 46% of the whole time of servicing a request.

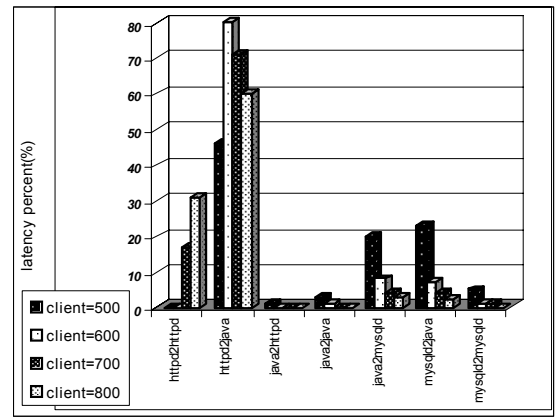

Fig.15. The latency percentages of components

At the same time, the latency percentage of httpd2httpd (first tier) increases dramatically from 17% (700 clients) to 31% (800 clients). We observe the CPU usage of the JBoss node is less than 60% and the I/O usage rate is not high. When servicing a request, the httpd2httpd is before the httpd2java in a causal path. So we can confirm that there is something wrong with the interaction between the httpd and the JBoss. Through reading the manual reference of RUBiS, we confirm that the problem may be mostly related with the configuration of thread pool in the JBoss. According to the manual book of the JBoss, one parameter named MaxThreads controls the max available thread number, and one thread services a connection. The default value of MaxThreads is 40.

We set MaxThreads as 250 and run the experiments again. In Fig.16, we observe that our work is effective. When concurrent clients increase from 500 to 800, the throughput is increased and the average response time is decreased in comparison with that of the default configuration. However, for 900 concurrent clients, the resource limit of hardware platform results in the new bottleneck. In Fig.16, TP_MT40 is the throughput when MaxThreads is 40, and RT_MT250 is the average response time when MaxThreads is 250.

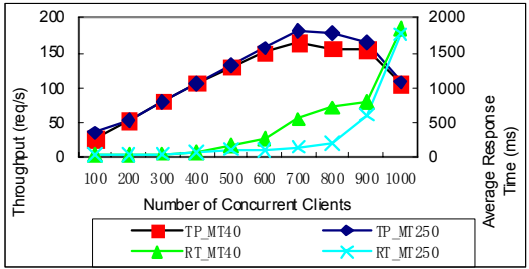

Fig.16. Performance for different MaxThreads.

**5.4.2. Injected performance problem.** To further validate the accuracy of locating performance problems using PreciseTracer, we have injected several performance problems into RUBiS components and the host nodes: for abnormal case 1, we modify the RUBiS code to inject a random delay into the second tier; for abnormal case 2, we lock the items table of the RUBiS database to inject a delay into the third layer; for abnormal case 3, we change the configuration of the Ethernet driver from 100 M to 10 M on the node running the JBoss.

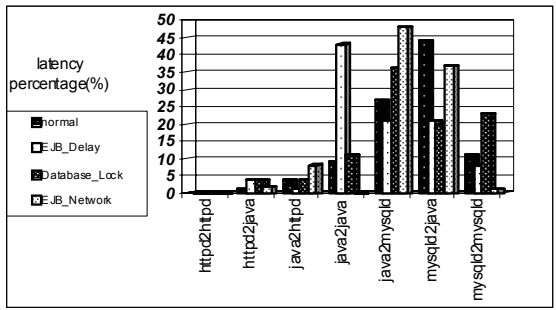

Fig.17. Latency percentages of components for abnormal cases

We use PreciseTracer to locate the component in question where different performance problems are injected. Fig.17 shows the latency percentages of components for normal case and three abnormal cases.

For abnormal case 1 (EJB_Delay), the latency percentage of Java2Java (JBoss, second tier) increases from less than 10% for the normal case to more than 40%, and the latency percentages of other components decrease with different amounts. So we can confirm that JBoss is in question.

For abnormal case 2 (DataBase_Lock), the latency percentage of mysqld2mysqld (third tier) increases from 12% for the normal case to more than 20%, and the latency percentage of java2mysqld (interaction from second tier to third tier) increases from 26% for the normal case to more than 35%. The Latency percentages of other components keep unchanged or decrease. So we confirm that MySQL is in question.

For the abnormal case 3 (EJB_Network), the latency percentage of Java2mysqld (from second tier to third tier) increases from 26% for the normal case to 47%; mysqld2java (from third tier to second tier) keeps about 37%. The latency percentage of httpd2java from first tier to second tier increases from 1% to 2%; the percentage of java2httpd from second tier to first tier increases from 4% to 8%. We observe that most of time of servicing request is spent on the interactions between second tier and third tier, and the three latency percentages of four interactions related with *the second tier* are increased. We confirm the second tier is in question. Further observation shows the latency percentage of Java2java strangely decreases from 9% to almost 0%. So we confirm that there is something wrong with the network of second tier.

## 6. Related work

### 6.1. Black box model

A much earlier project, DPM, [11] instruments the operating system kernel and tracks the causality between pairs of message to trace unmodified applications. DPM is not precise, since the existence of a path in the resulting graph does not necessarily mean that any real causal path followed all of those edges in that sequence [3].

Project5 [3] and WAP5 [2] accept the imprecision of probabilistic correlation methods. Project5 [3] uses the time series analysis to infer causal paths in a distributed system of black boxes from the relative timestamps of network traffic. More recently WAP5 [2] infers causal paths for wide-area systems from tracing stream on a per-process granularity using library interposition. The E2Eprof [6] proposes a pathmap algorithm, similar to the convolution algorithm of Project5, but uses compact trace representations and a series of optimizations make it suitable for online performance diagnosis.

Using the knowledge of protocols, BorderPatrol [4] isolates and schedules event or request at the protocol level to precisely trace requests. BorderPatrol requires writing many protocol processors and requires more specialized knowledge than pure black-box approach.

### 6.2. Non-black box model

The most invasive systems, such as Netlogger [12] and ETE [13] require programmers to add event logging to carefully-chosen points to find causal paths rather than infer them from passive traces. Pip [15] inserts annotations into the source code to log actual system behavior, but can extract causal path information with no false positives or false negatives. Magpie [1] collects events at different points in a system and uses an event schema, which is application-specific, to correlate these events into causal paths. Stardust [14], a system used as an on-line monitoring tool in a distributed storage system, is implemented in a similar way. Whodunit [16] annotates profile data with transaction context synopsis and provides finer grained knowledge of transactions within each box.

To avoid the modification of applications' source code, some work enforces middleware or infrastructure changes. Pinpoint [17] locates component faults in J2EE platform by tagging and propagating a globally unique request ID with each request. Causeway [18] enforces change to network protocol so as to tag meta-data with existing module communication. X-Trace [19] modifies each network layer to carry X-Trace meta-data that enables path casual path reconstruction and focuses on debugging paths through many network layers.

## 7. Conclusion

We have developed the PreciseTracer tool to help users understand and debug performance problems of a multi-tier service of black boxes. Our contributions are two-fold: (1) we have designed a precise tracing algorithm to derive causal paths for each individual request, using interaction activities of components of black boxes. Our algorithm only uses the *application-independent knowledge*, such as timestamps and end-to-end communication channels, which are available from the operating system; (2) we have presented a component activity graph (CAG) abstraction to represent causal paths of requests and facilitate end-to-end performance debugging of a multi-tier service. Our experiments show that one can successfully pinpoint performance problems from observed changes of latency percentages of components, calculated from CAGs.

In the near future, we will propose the mathematical foundation for automatic performance debugging.

## 8. Acknowledgements

We are very grateful to anonymous reviewers, the DSN shepherd and Prof. Weisong Shi for their helpful suggestions. This paper is supported by the NSFC (Grant No. 60703020) and the 863 programs (Grant No. 2009AA01Z139 and Grant No. 2006AA01A102).